\documentclass[prd,twocolumn,superscriptaddress, nofootinbib, preprintnumbers]{revtex4-1}
\usepackage{amsmath}
\usepackage{graphicx}
\usepackage{xcolor}
\usepackage{hyperref}
\usepackage{float}
\usepackage{braket}
\usepackage{comment}
\usepackage{ulem}

\newcommand{\be}{\begin{equation}}
\newcommand{\ee}{\end{equation}}

\begin{document}
\preprint{YITP-SB-2021-02}
\title{Neutrinos in N-body simulations}
\author{Caio Bastos de Senna Nascimento and Marilena Loverde \\{\it{\small  C.N. Yang Institute for Theoretical Physics, Department of Physics \& Astronomy, Stony Brook University, Stony Brook, NY 11794}}}
\begin{abstract}
In the next decade, cosmological surveys will have the statistical power to detect the absolute neutrino mass scale. N-body simulations of large-scale structure formation play a central role in interpreting data from such surveys. Yet these simulations are Newtonian in nature. We provide a quantitative study of the limitations to treating neutrinos, implemented as N-body particles, in N-body codes, focusing on the error introduced by neglecting special relativistic effects. Special relativistic effects are potentially important due to the large thermal velocities of neutrino particles in the simulation box. We derive a self-consistent theory of linear perturbations in Newtonian and non-relativistic neutrinos and use this to demonstrate that N-body simulations overestimate the neutrino free-streaming scale, and cause errors in the matter power spectrum that depend on the initial redshift of the simulations. For $z_{i} \lesssim 100$, and neutrino masses within the currently allowed range, this error is $\lesssim 0.5\%$, though represents an up to $\sim 10\%$ correction to the shape of the neutrino-induced suppression to the cold dark matter power spectrum.  We argue that the simulations accurately model non-linear clustering of neutrinos so that the error is confined to linear scales.
\end{abstract}
\maketitle

\section*{Introduction}
\label{sec:intro}

Neutrino oscillation experiments have established that at least two of the neutrino mass eigenstates have a non-zero mass, giving a lower bound on the sum of all three masses as $\sum m_{\nu} \gtrsim 0.06$ eV, $0.1$ eV for the normal and inverted hierarchies, respectively (see, e.g. \cite{1708.01186, 1806.11051}). At present, it is very hard to determine the neutrino mass scale at these lower limits with laboratory experiments. The current constraints from the KATRIN experiment, based on measurements of $\beta$-decay, are $m_{\nu,i} < 1.1$ eV at $90\%$ confidence \cite{Aker:2019uuj}. Tighter constraints on the upper bound to the sum of neutrino mass eigenvalues come from cosmological datasets, providing $ \sum m_{\nu} \lesssim (0.11-0.24)$ eV \cite{Stocker:2020nsx, 1811.02578, 1807.06209}, depending on the specific choice of data. Furthermore, cosmological observations point to a universe that is today dominated by dark energy and dark matter: all the evidence suggests the presence of physics beyond the standard model (see, for instance, \cite{1907.12409} for a review).  

Current and future large-scale structure surveys \cite{1907.08945, 1910.09273, 1906.10134, 1305.5422, 1508.04473, 1111.6398, Levi:2019ggs} will shed some light on the nature of this, as of now, mysterious new physics. Concretely, we expect for example, to be able to determine the absolute mass scale of neutrinos (and potentially the hierarchy) \cite{1903.03689, 1808.05955, 0907.1917}, further constrain the dark energy equation of state \cite{Slosar:2019flp}, and number of effective neutrino species \cite{Green:2019glg}. Large-scale structure surveys will probe small scales where nonlinear gravitational evolution is important. This adds one additional degree of complexity to the mission, as we need to rely on simulations of nonlinear structure formation, typically executed by running N-body codes such as \cite{1609.08621, 1003.2422, 1611.01545, 2007.13394, 1505.07148, 1301.0322, 2011.12504}, to accurately interpret the cosmological observables. The simulations are, however, Newtonian in nature, though large-scale general relativistic (GR) effects can be fully recovered in the case of a dark-matter-only universe \cite{1101.3555, 1210.5446, 1505.04756}. Developing a fully-relativistic N-body code is an active area of research \cite{1904.07841, 1905.08890, 1707.06938}.

In this work we will consider limitations to treating neutrinos, implemented as N-body particles, in Newtonian codes. Our analysis relies solely on individual particles evolving with the Newtonian equation of motion. An example of this in the literature would be the codes described in \cite{1801.03906,1109.4416, 1311.0866, 1708.01154, 2007.15279}, all based on GADGET-3, which is itself an improved version of GADGET-2 \cite{2006.10203, astro-ph/0505010}. The discussion in this paper is restricted to simulations that treat neutrinos as N-body particles, as opposed to a hydrodynamic approach, e.g. \cite{2002.04601, 1003.2422}.

We will be interested in special relativistic (SR) effects that are neglected in a Newtonian treatment of neutrinos. To illustrate the necessity of considering SR effects for neutrinos, consider an individual neutrino of mass $m_{\nu} = 0.05$eV, the minimum value required for at least one state by neutrino oscillation data, and a value just below that set by current limits on the sum of neutrino mass states \cite{1806.11051, Stocker:2020nsx}. For a simulation starting at redshift of $z_{i} \approx 100$, the mass-to-temperature ratio at the initial redshift is $m_{\nu}/(1+z_{i})T_{\nu,0} \approx 3$, where $T_{\nu,0} \approx 1.95K \approx 1.7 \times 10^{-4}$eV is the temperature of relic neutrinos today. In this circumstance, one should worry that evolving neutrinos with the Newtonian limit of the geodesic equation, equivalent to suppressing both GR and order $(v/c)^2$ corrections (what we call, from now on, Newtonian neutrinos), can introduce significant systematic error in the output of N-body simulations. 

As we shall see, particles with large thermal velocities evolved with non-relativistic equations of motion will travel faster and further than they would if evolved with the correct, relativistic equations of motion. Consequently, N-body simulations with neutrinos will overestimate the neutrino free-streaming scale, $\lambda_{\textrm{fs}} \sim v_{\nu}/(aH)$, where $v_\nu$ is the neutrino velocity dispersion and $H$ is the Hubble rate. This overestimation disappears at late times, as $z \to 0$, but the neutrino horizon, $\lambda_{\textrm{h}} = \int_{z_{i}} d(\ln a) \lambda_{\textrm{fs}}$, which depends on the entire history of the free-streaming scale, will be significantly overestimated, even as $z \to 0$. This shift on the free-streaming scale could have a perceptible impact on the usual neutrino-induced suppression to the growth of structure \cite{astro-ph/0603494}.

To study the impact of these errors on the evolution of matter perturbations, we develop the exact linear-theory evolution of inhomogeneities in the distribution of Newtonian neutrinos. We derive the usual fluid-approximation for Newtonian neutrinos (FA) achieved by introducing an effective sound speed, as well as leading-order corrections that include the effects of a non-zero shear stress. For $z \lesssim 100$, these effects impact only the small-scale oscillations in the neutrino transfer functions, and can hence be safely neglected when obtaining matter perturbations. The FA can be used to generate initial conditions for N-body simulations \cite{1605.05283}, through rescaling of the power spectrum from $z=0$ to the simulation initial redshift $z_{i}$.

The FA is missing both SR and GR corrections. In order to isolate the SR effects, which are only important for fast moving particles such as neutrinos, we also develop a non-relativistic fluid approximation that ignores only SR terms that are missing for neutrinos, but keeps GR terms, and therefore produces the correct evolution of both cold dark matter and non-relativistic neutrino perturbations on large-scales, by default. By comparing our non-relativistic fluid approximation to the correct fully-relativistic one, we quantify the errors in the usual neutrino-induced suppression to the growth of structure, caused by treating neutrinos as non-relativistic particles. 

In Newtonian simulations of a dark-matter-only universe, these GR effects we are including can be obtained with an adjustment of initial and final displacements of particles \cite{1101.3555, 1210.5446}. In \cite{1505.04756}, it was realized that this is no coincidence: As long as the output of the simulations is understood in terms of the so-called N-body gauge, GR effects can be fully recovered, i.e. they can be accommodated into a gauge transformation (also see \cite{1606.05588}). In this scenario, one can further include the effect of light neutrinos and photons, assuming a linear evolution for these components \cite{1811.00904, 1811.12412, 2003.07387}. The N-body gauge approach can also be generalized to cosmologies with massive neutrinos, where the CDM+$\nu$ center-of-mass motion is followed by N-body simulations \cite{1807.03701}.  Relativistic corrections can also be studied in beyond $\Lambda$CDM scenarios \cite{2006.11019}.

This paper is organized as follows. In \ref{sec:NewtonianneutrinosBG}, we review the Newtonian limit of the geodesic equation in an isotropic universe, and present expressions for the background energy density and pressure of both Newtonian and relativistic relic neutrinos. In \ref{sec:linear}, we present self-consistent linear perturbation theories of neutrino and cold dark matter (CDM) particles evolved according to the Newtonian and non-relativistic ($(v/c)^2 \ll 1$) equations of motion.  In \ref{sec:numericalresults}, we present numerical results using the linear theories from \ref{sec:linear}, to compare the CDM transfer functions in the presence of non-relativistic neutrinos, to the same quantities evolved with the fully-relativistic equations of motion. We also compare the linear theories from \ref{sec:linear} with the cosmic linear anisotropy solving system (CLASS), in order to illustrate the additional sources of systematic error, in N-body simulations, that are not being accounted for in our analysis (such as radiation). In \ref{sec:implications} we discuss the implications of our results for N-body simulations with neutrinos implemented as particles, and in \ref{sec:conclusions} we summarize our findings. In an appendix \ref{sec:fluidderiv}, we present the complete derivation of fluid equations for  Newtonian and non-relativistic neutrino components. While our study is motivated by neutrinos co-evolving with cold dark matter (CDM) and baryons (b), the analysis makes no assumptions about the particle nature of either component, only that they are non-interacting. In what follows we may therefore refer to the cold non-relativistic component as dust and the non-cold dark matter component as NCDM (following the notation of CLASS \cite{1104.2935}).

\section{Dynamics of Newtonian Neutrinos: Background}
\label{sec:NewtonianneutrinosBG}

Let us start our analysis at the level of background. The relativistic dynamics follows from the geodesic equation in a Friedmann-Robertson-Walker universe with line element
\begin{align}
ds^2 &= -dt^2 + a^2(t)d\vec{x}^2\,,\\
&=a^2(\tau)\left(-d\tau^2 +d\vec{x}^2\right)\,.
\end{align}
The geodesic equation is given by
\be
\label{eq:geodesic}
\frac{d\vec{v}}{dt} + \Big(H+ \frac{1}{\gamma} \frac{d \gamma}{dt}\Big)\vec{v} = 0\,,
\ee
where $\vec{v} = ad\vec{x}/dt$ is the peculiar velocity, $\vec{x}$ is the comoving position vector, $1/\gamma^2 = 1-v^2$,  and $H=a^{-1}da/dt$ is the Hubble rate. The evolution of a Newtonian component arises from the assumption $H \gg \gamma^{-1} d \gamma/dt = \gamma^2 \vec{v} \cdot d\vec{v}/dt $, and reads,
\be
\label{eq:Newevolution}
\frac{d\vec{v}}{dt} + H\vec{v} = 0\,.
\ee
This implies $v \propto a^{-1}$. In contrast, the correct equation of motion for a relativistic particle, Eq.~(\ref{eq:geodesic}), will produce $p = \gamma m v \propto a^{-1}$, where $p$ is the momentum, so that the evolution of $v$ is more complicated.

\begin{figure*}
		\centering
		\includegraphics[width=1\textwidth]{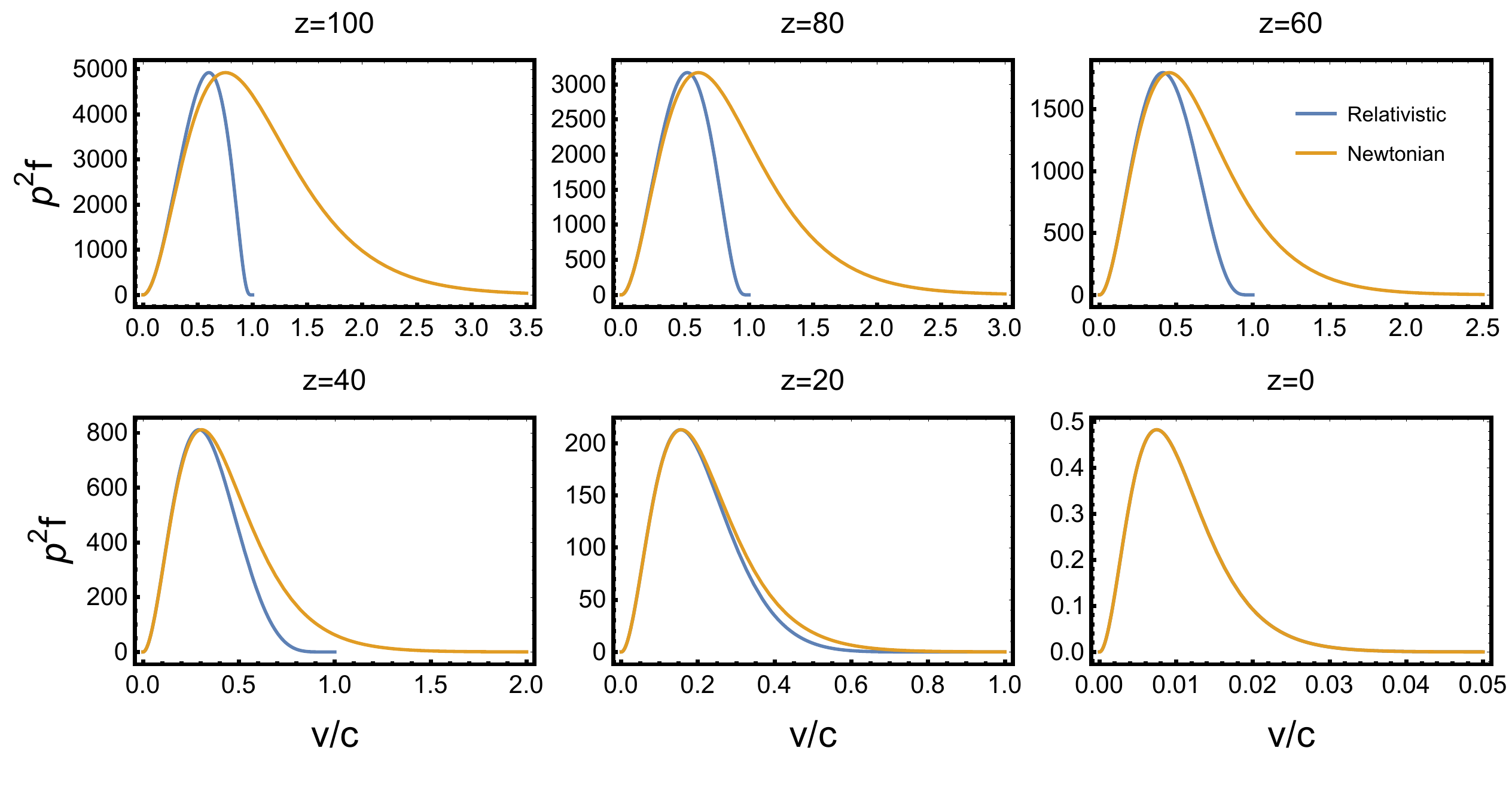}

		\caption{Evolution of the velocity distribution function, for $m_{\nu}=0.05$eV and varying redshift. A significant portion of particles, in the simulation box, are superluminal for $z\gtrsim 50$. The differences in the velocity distribution function shown above do not depend on $z_i$, the initial redshift of the simulations. } \label{fig:velocitydist005}
\end{figure*}

 In Newtonian N-body simulations, neutrinos are given a thermal velocity that is obtained by sampling from a Fermi-Dirac distribution,
\be
\label{eq:FD}
f_{0}(p) = \frac{2}{(2\pi)^3} \frac{1}{e^{\frac{p}{T_i}}+1}\,,
\ee 
at the initial redshift $z_{i}$, where $p=mv$ is the nonrelativistic expression for the physical momentum, and $T_{\nu,i} =(1+z_{i})T_{\nu,0}$ is the neutrino temperature at $z_i$ \footnote{Note that in converting the exact momentum distribution in  Eq.~(\ref{eq:FD}) into a velocity distribution function at $z_i$, one could use $p=mv$, as we have assumed, or solve for $v$ in the equation $p = \gamma m v$. We assume the choice $p=mv$, since it is the one made in (at least some) Newtonian codes, e.g. \cite{1909.05273}. As shown in Figure \ref{fig:velocitydist005}, this leads to a velocity distribution function that deviates from its relativistic counterpart at the initial redshift, while agreeing with the latter today, i.e. as $z \to 0$. If the choice $p = \gamma m v$ was made, then there would be no difference at the initial redshift, by construction, at the expense of getting the $z \to 0$ limit wrong, given the Newtonian evolution. }. Then, under the Newtonian dynamics $p= mv \propto a^{-1}$, as it would if the correct relativistic expression were used, and the distribution retains the form of a Fermi-Dirac distribution, with a temperature that scales as $T_{\nu} \propto a^{-1}$. 

\begin{figure*}
		\centering
		\includegraphics[width=1\textwidth]{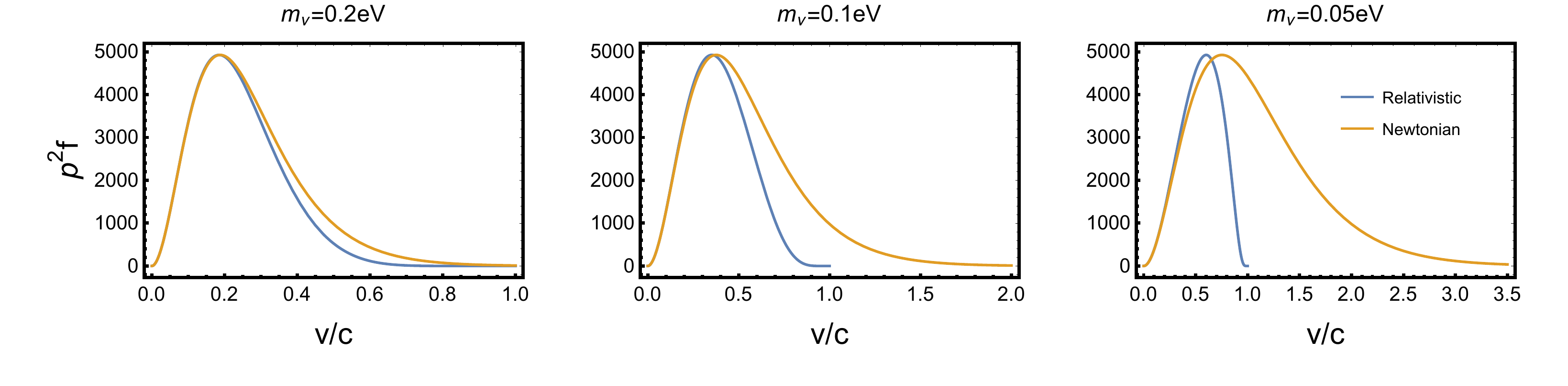}
		\caption{The velocity distribution function at $z=100$ for different values of $m_\nu$. The two distributions approach one another for $m_{\nu} \gtrsim 0.2eV$.} \label{fig:velocitydist02}
\end{figure*}

In Figures \ref{fig:velocitydist005} and \ref{fig:velocitydist02} we plot the velocity distribution function, and compare to its fully relativistic counterpart where $p=\gamma mv$ is used to define the physical momentum. A few comments are in order. First notice that, for a neutrino mass of $m_{\nu} = 0.05$eV and $z\gtrsim 50$, a significant portion of the particles in the simulation box will be superluminal. This happens at any redshift for which $m_{\nu}/3T_{\nu} \lesssim 2$. On the other hand, for $m_{\nu} \gtrsim 0.2$eV, the difference in the distributions is rather small already at $z=100$, so we expect the Newtonian limit of the geodesic equation to be a good approximation. This remains true for $m_{\nu}/3T_{\nu} \gtrsim 4$. As we shall see, the difference in the velocity distribution functions, shown in Figures \ref{fig:velocitydist005} and \ref{fig:velocitydist02}, can potentially introduce errors in both linear and nonlinear clustering of neutrinos.

It is also clear, from Figures  \ref{fig:velocitydist005} and \ref{fig:velocitydist02}, that the velocity dispersion $v_{\nu}$ of Newtonian neutrinos,
\be 
	v_{\nu, \textrm{New}}^2 = \frac{\int_{0}^{\infty} dq q^2f_{0}(q) \Big(\frac{q}{ma}\Big)^2}{\int_{0}^{\infty} dq q^2f_{0}(q)}
\ee
will be bigger than that of relativistic neutrinos,
\be 
	v_{\nu, \textrm{rel}}^2 = \frac{\int_{0}^{\infty} dq q^2f_{0}(q) \Big(\frac{q}{ma}\Big)^2 \frac{1}{1+\big(\frac{q}{ma}\big)^2}}{\int_{0}^{\infty} dq q^2f_{0}(q)}
\ee

We now consider a fluid description of both Newtonian and relativistic neutrinos. For three relativistic neutrinos of equal mass, the energy density and pressure are given by,
\begin{align}
\label{eq:densityrel}
\rho_{\textrm{rel}} &= 4\pi N_{\nu} ma^{-3} \int_{0}^{\infty} dq q^2f_{0}(q) \sqrt{1+\big(\frac{q}{ma}\big)^2}\,, \\ 
\label{eq:pressurerel}
P_{\textrm{rel}} &= \frac{4\pi N_{\nu}}{3}ma^{-3} \int_{0}^{\infty} dq q^2f_{0}(q)  \Big(\frac{q}{ma}\Big)^2 \frac{1}{\sqrt{1+\big(\frac{q}{ma}\big)^2}}\,,
\end{align}
where $q=pa$, and $N_{\nu}=3$ is the degeneracy factor. For a Newtonian component we have, 
\begin{align}
\label{eq:densitynew}
\rho_{\textrm{New}} &= 4\pi N_{\nu} m a^{-3} \int_{0}^{\infty} dq q^2f_{0}(q)\,, \\ 
\label{eq:pressurenew}
P_{\textrm{New}} &= \frac{4\pi N_{\nu}}{3} ma^{-3} \int_{0}^{\infty} dq q^2f_{0}(q) \Big(\frac{q}{ma}\Big)^2\,.
\end{align}

\begin{figure}
		\centering
		\includegraphics[width=0.45\textwidth]{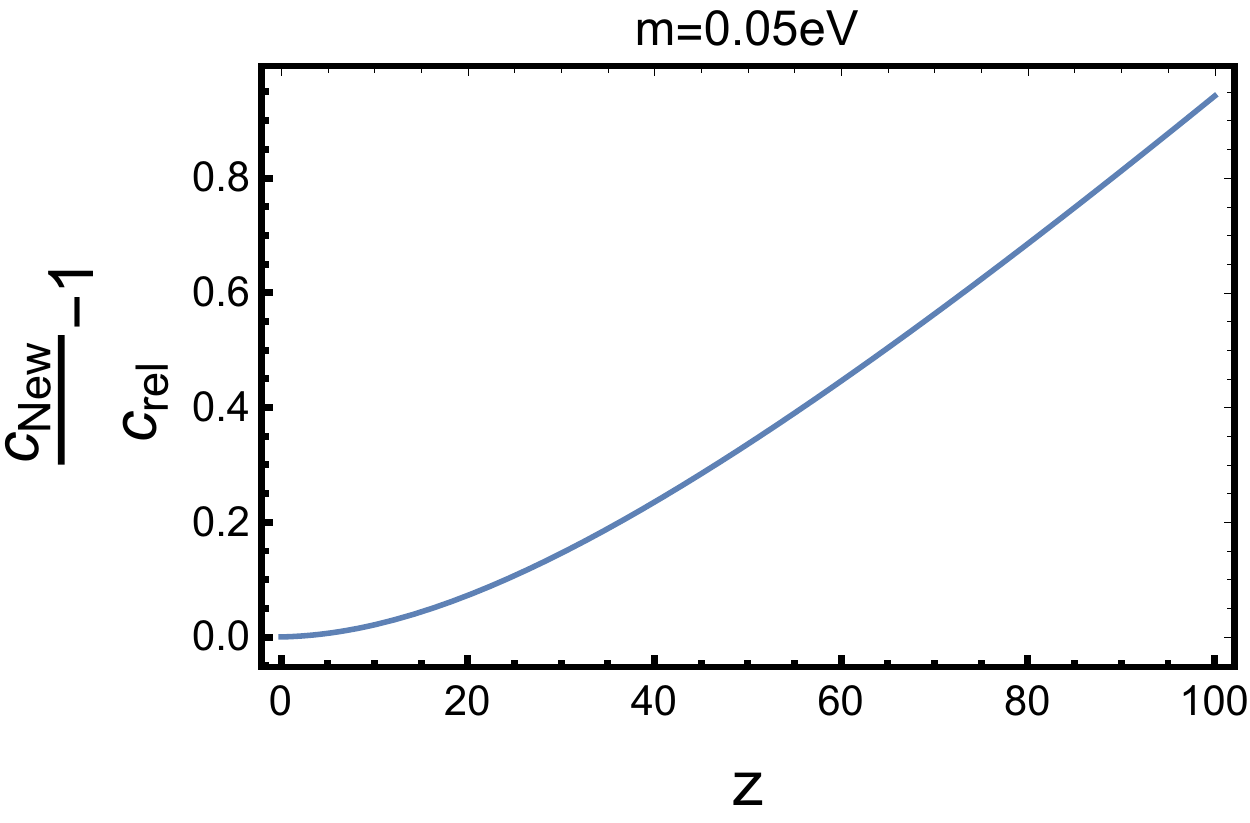}
		\caption{Relative difference between Newtonian and relativistic adiabatic sound speeds for a neutrino mass of $m_{\nu} = 0.05$eV.} \label{fig:cs}
\end{figure}

In order to define the free-streaming scale, we also introduce the adiabatic sound speed,
\be
\label{eq:cg}
c_g^2 = \frac{\dot{P}}{\dot{\rho}}\,,
\ee
where $\dot{P}$ is the time-derivative of the neutrino fluid pressure, and $\dot{\rho}$ the time-derivative of the energy density. The relative difference in the Newtonian and relativistic adiabatic sound speeds is presented in Figure \ref{fig:cs}. From this it follows at once that the instantaneous free-streaming scale \footnote{In order to define the free-streaming scale, the (adiabatic) sound speed should be used instead of the velocity dispersion. Note that $c_{g, \textrm{New}} = \frac{\sqrt{5}}{3} v_{\nu, \textrm{New}}$, while $c_{g, \textrm{rel}} \approx \frac{\sqrt{5}}{3} v_{\nu, \textrm{rel}}$ in the non-relativistic regime (see e.g. the appendix A of \cite{1003.0942}).}
\be
\lambda_{\textrm{fs}} \equiv  \frac{c_{g}}{aH}\,,
\ee 
is overestimated for Newtonian neutrinos and therefore $k_{\textrm{fs}} = 2\pi/\lambda_{\textrm{fs}}$ is underestimated in the simulations. Since $c_{g,\textrm{New}} \to c_{g,\textrm{rel}}$ as $z \to 0$, the free-streaming is correctly reproduced at late times. This difference in free-streaming scales is important, as $\lambda_{\textrm{fs}}$ defines the length scale below which neutrino perturbations get completely washed out. Also relevant is the neutrino (fluid) horizon,
\be
\lambda_{\textrm{h}} = \int_{z_{i}} \frac{dt}{a}\, c_{g}(a) = \int_{z_{i}} d(\log a) \lambda_{\textrm{fs}} \,,
\ee
or in k-space $k_{\textrm{h}} = 2\pi/ \lambda_{\textrm{h}}$. This integral is dominated by its lower limit, where Newtonian and relativistic free-streaming scales are most different, so the underestimation in the horizon scale persists in the limit $z \to 0$. This defines the length scale above which free-streaming is no longer relevant, with the neutrino component behaving like cold matter. See \cite{1803.11545,1404.1740} for short reviews on the effect of neutrinos in cosmology. 

It is clear from the above that the Newtonian neutrino energy density will not have the correct evolution with time. This can potentially impact the evolution of the Hubble rate as a function of time through the Friedmann equation as well. But,  in N-body simulations, the background energy densities, and therefore the Hubble rate, are not computed from the dynamics of N-body particles. Moreover, one can straightforwardly provide a tabulated set of values for $H(a)$ computed from the correct relativistic dynamics \cite{1605.05283, 1209.0461}, i.e.
\be
\Big(\frac{H}{H_{0}}\Big)^2 = \Omega_{\textrm{r},0}a^{-4} + \Omega_{\textrm{d},0}a^{-3} + \Omega_{\Lambda,0} + \frac{\rho_{\textrm{rel}}(a)}{3M_{\textrm{pl}}^2H_{0}^2}\,.
\ee
where $M_{\textrm{pl}}^2 = 1/8\pi G$, $H_0=100h \,{\rm km/s/Mpc}$ is the Hubble parameter today, $\Omega_{\textrm{r},0}$ is the radiation density today, $ \Omega_{\textrm{d},0}$ is the energy density in cold dark matter and baryons today, $\Omega_{\Lambda,0}$ is the energy density in vacuum, and $\rho_{\textrm{rel}}$ is the relativistic energy density of the relic neutrinos, as computed from Eq.~(\ref{eq:densityrel}). 

The differences in the Newtonian and correct relativistic neutrino velocity dispersions, established in Figures \ref{fig:velocitydist005}-\ref{fig:cs}, will impact the evolution of neutrino perturbations in N-body simulations. This could limit the accuracy to which N-body simulations can model neutrinos, at both the linear and non-linear regimes.

\section{Dynamics of Newtonian Neutrinos: Linear Perturbations}
\label{sec:linear}
In this section we will present a self-consistent theory of linear perturbations of neutrinos and dust evolved according to non-relativistic equations of motion. As we shall see, this will amount to taking two independent limits of the fully relativistic equations for linear perturbations. First, as in the last section, we will take the limit $(v/c)^2 \ll 1$ in the equation of motion for neutrinos, as well as for sources to the Poisson equation. Second, to make contact with N-body simulations and the formalism used to generate their initial conditions, we present in Sec. \ref{subsec:Nbodyfluid} evolution equations ignoring time derivatives of metric perturbations that appear in the linearized geodesic equation, but are not included in N-body codes. These dropped terms are referred to as {\it GR-terms}, and their particular form depends on the choice of gauge for the perturbed metric. In the N-body gauge they disappear entirely for a universe comprised of only dust and $\Lambda$, but cannot be completely eliminated in the presence of relativistic particles \cite{1505.04756}. Since our primary interest is to study the impact of SR effects ($v/c \sim 1$), we will work in the Newtonian gauge, where the interpretation of variables is more straightforward, and present final results that include the GR terms. 

\subsection{Framework}
The full details of our calculations of linear perturbations of neutrinos and dust are left to Appendix \ref{sec:fluidderiv}, but we shall outline our approach here before proceeding. To study the evolution of linear perturbations in dust and neutrinos we start from the collisionless Boltzmann Equation\footnote{For dust, of course, one does not need the full Boltzmann equation and can start directly with the equations for a pressureless, shearless fluid.}, 
\be
\label{eq:Boltzmanndef}
	\frac{df}{d\tau} = \frac{\partial f}{\partial \tau} + \frac{dx^{i}}{d\tau} \frac{\partial f}{\partial x^{i}} + \frac{dq^i}{d\tau} \frac{\partial f}{\partial q^{i}} = 0\,,
\ee
which governs the dynamics of the distribution function in phase space. In the above, $\vec{q} = a \vec{p}$, with $\vec{p}$ the proper momentum and $\tau$ the conformal time. The distinction between neutrinos and dust evolved as Newtonian particles, as is done in N-body codes, or as particles subject to non-relativistic or fully-relativistic dynamics, amounts to implementing different definitions of particle momentum and equations of motion (that is, different expressions for $ dq^i/d\tau$ and $dx^{i}/d\tau$ in Eq.~(\ref{eq:Boltzmanndef})). These differences will lead to different evolution equations for $f(x, q, \tau)$ that capture how a distribution of particles would evolve, subject to Newtonian, non-relativistic, or relativistic dynamics. 

In the following subsections we will use this approach to derive several different systems of equations for the linear evolution of neutrinos and dust. While it is a bit cumbersome, we will use subscripts to distinguish between the quantities that satisfy the different evolution equations. We will first consider strict Newtonian evolution, as implemented in N-body codes, these quantities being identified by the subscript $_{\textrm{N-body}}$. Subsequently, we will impose a fluid approximation to the Newtonian $_{\textrm{N-body}}$ equations, and these quantities are denoted by $_{\textrm{FA}}$. In Sec. \ref{subsec:NRF}, we present a fluid approximation that ignores SR terms for neutrinos, but keeps GR terms for both neutrinos and dust. We refer to this as the non-relativistic fluid approximation, and identify variables solving those equations with $_{\textrm{NRF}}$. Finally, we show the usual CLASS relativistic fluid approximation for the coevolution of neutrinos and dust in Sec. \ref{subsec:RF}. The variables solving these equations will be identified by a subscript $_{\textrm{RFE}}$. Table \ref{table:t1} shows a list of all subscripts with their underlying assumptions.

\begin{table}[]
\begin{tabular}{|l|l|}
\hline
{\bf Subscript}& {\bf Assumptions}\\
\hline
N-body & Newtonian evolution                                                                                    \\ \hline
FA     & \begin{tabular}[c]{@{}l@{}}Newtonian evolution \\ + fluid approximation\end{tabular}                   \\ \hline
NRF    & \begin{tabular}[c]{@{}l@{}}Newtonian evolution\\ + GR corrections\\ + fluid approximation\end{tabular} \\ \hline
RFE   & \begin{tabular}[c]{@{}l@{}}Relativistic evolution\\ + fluid approximation\end{tabular}                 \\ \hline
\end{tabular}
\caption{Different subscripts with their associated underlying assumptions. Note that Newtonian evolution $+$ GR corrections is equivalent to dropping the SR terms (taking the $(v/c)^2\ll 1$ limit) of the fully relativistic expressions.}
\label{table:t1}
\end{table}

\subsection{Linear Perturbations in N-body Fluids }
\label{subsec:Nbodyfluid}

In N-body simulations particles evolve according to the Newtonian equation of motion, 
\be
\label{eq:NewEOM}
\frac{d\vec{v}_{\textrm{N-body}}}{dt} + H\vec{v}_{\textrm{N-body}} = -\frac{1}{a} \vec{\nabla} \psi_{\textrm{N-body}}\
\ee
where 
\be
\vec{v}_{\textrm{N-body}} = a \frac{d\vec{x}}{dt}
\ee
is the peculiar velocity, and $\psi_{\textrm{N-body}}$ is the gravitational potential, sourced only by the rest mass of the particles in the simulation, 
\be
\label{eq:Newpoissongeneral}
	k^2 \psi_{\textrm{N-body}} = - \frac{1}{2 M_{\textrm{pl}}^{2}} a^2 \sum_{i} \delta\rho_{\textrm{N-body}, i}
\ee
where $ \delta\rho_{\textrm{N-body}, i}$ is the perturbation to the mass density of species $i$, further ignoring the inhomogeneities in the local volume. As shown in Appendix \ref{sec:fluidderiv}, one can derive exact fluid equations for a system of non-interacting particles evolving according to Eq.~(\ref{eq:NewEOM}) and Eq.~(\ref{eq:Newpoissongeneral}). 

For Newtonian neutrinos in N-body simulations we have, 
\begin{subequations}
\label{eq:nufluidexact}
\begin{align}
\delta'_{\nu, \textrm{N-body}} &= -\theta_{\nu, \textrm{N-body}} \,,\\ 
\theta'_{\nu, \textrm{N-body}} &= -\mathcal{H} \theta_{\nu, \textrm{N-body}} + \frac{\delta P_{\nu,\textrm{N-body}}}{\delta \rho_{\nu,\textrm{N-body}}} k^2 \delta_{\nu, \textrm{N-body}}\\
& -k^2 \sigma_{\nu, \textrm{N-body}} + k^2 \psi_{\textrm{N-body}}\,,\nonumber
\end{align}
\end{subequations}
along with other evolution equations for higher order multipole moments of the perturbed distribution function. Here $' = d/d\tau$, $\delta_{\nu,\textrm{N-body}} = \delta \rho_{\nu,\textrm{N-body}}/\rho_{\nu,\textrm{New}}$, with $\delta\rho_{\nu,\textrm{N-body}}$ the perturbation to the neutrino mass density, $\theta_{\nu,\textrm{N-body}}$ is the velocity divergence, $\delta P_{\nu,\textrm{N-body}}$ the perturbation to the pressure, $\sigma_{\nu,\textrm{N-body}}$ the anisotropic stress. 

The dust component evolves according to the same equations in the limit of no anistropic stress and no pressure, 
\begin{subequations}
\label{eq:dustNbody}
\begin{align}
\delta'_{d, \textrm{N-body}} &= -\theta_{d, \textrm{N-body}}  \,,\\
\theta'_{d, \textrm{N-body}} &= -\mathcal{H} \theta_{d, \textrm{N-body}}  + k^2 \psi_{\textrm{N-body}}\,.
\end{align}
\end{subequations}
In Eq.~(\ref{eq:nufluidexact}) and Eq.~(\ref{eq:dustNbody}), $\psi_{\textrm{N-body}}$ is the potential computed from the Poisson equation sourced by the N-body fluids,

\be
\label{eq:Newpoisson}
	k^2 \psi_{\textrm{N-body}} = - \frac{1}{2 M_{\textrm{pl}}^{2}} a^2 (\rho_{d} \delta_{d, \textrm{N-body}} + \rho_{\nu, \textrm{New}} \delta_{\nu, \textrm{N-body}})\,.
\ee
Notice that this is sourced by the Newtonian expression for the neutrino energy density, Eq.(\ref{eq:densitynew}), i.e. it just includes the rest mass of neutrino particles. 

As described in Appendix \ref{sec:fluidderiv}, each of these quantities is computed self-consistently assuming a Newtonian treatment of each of the neutrino properties. This system of equations should therefore describe the exact linear evolution of neutrinos and dust within N-body simulations. To proceed, of course, one needs additional expressions for $\delta P_{\nu,\textrm{N-body}}/\delta \rho_{\nu,\textrm{N-body}}$ and $ \sigma_{\nu, \textrm{N-body}}$ to close the system of equations. 

\subsection{Two-fluid approximation (FA) for N-body neutrinos and dust }
\label{subsec:FA}
The standard two-fluid approximation for neutrinos and dust is obtained from the N-body equations, in Sec. \ref{subsec:Nbodyfluid}, by setting the anisotropic stress to zero, $\sigma_{\nu, \textrm{N-body}}\approx 0$, and ratio of the pressure and energy perturbations to the Newtonian adiabatic sound speed squared, $\delta P_{\nu,\textrm{N-body}}/\delta \rho_{\nu,\textrm{N-body}}\approx c_{g,\textrm{New}}^{2}$, obtained from Eqs.~(\ref{eq:densitynew}) and (\ref{eq:pressurenew}). This yields, 
\begin{subequations}
\label{eq:nuFA}
\begin{align}
\delta'_{\nu, \textrm{FA}} &= -\theta_{\nu, \textrm{FA}}\,, \\ 
\theta'_{\nu, \textrm{FA}} &= -\mathcal{H} \theta_{\nu, \textrm{FA}} +  c_{g,\textrm{New}}^{2} k^2 \delta_{\nu, \textrm{FA}} + k^2\psi_{\textrm{FA}}\,.
\end{align}
\end{subequations}
This is precisely the fluid approximation, used to generate initial conditions for N-body simulations, by rescaling of the matter power spectrum from $z=0$ to the initial simulation redshift $z_i$ \cite{1605.05283}. Initial particle positions and velocities are obtained with the Zeldovich approximation \cite{Zeldovich:1969sb}, or the more accurate higher-order Lagrangian perturbation theory schemes \cite{astro-ph/9406013, 2008.09588}. Note that our derived sound speed, as follows from Eq.~(\ref{eq:FD}), Eq.~(\ref{eq:cg}), Eq.~(\ref{eq:densitynew}), and Eq.~(\ref{eq:pressurenew}),
\be
c_{g,\textrm{New}}^{2} = \frac{25}{3} \frac{\zeta(5)}{\zeta(3)} \big(\frac{T_{0,\nu}}{m_{\nu}}\big)^2(1+z)^2 \approx 7.19\big(\frac{T_{0,\nu}}{m_{\nu}}\big)^2(1+z)^2
\ee
is the expression used in \cite{1605.05283, 1408.2995}. The dust component in Eq.~(\ref{eq:dustNbody}) is already treated like a presureless fluid, so the FA equations are identical to the N-body ones\,
\begin{subequations}
\label{eq:dFA}
\begin{align}
\delta'_{d, \textrm{FA}} &= -\theta_{d, \textrm{FA}}  \,,\\
\theta'_{d, \textrm{FA}} &= -\mathcal{H} \theta_{d, \textrm{FA}}  + k^2 \psi_{\textrm{FA}}\,.
\end{align}
\end{subequations}
Similarly, the only modification to the Poisson equation in Eq.~(\ref{eq:Newpoisson}) is to change the source terms to the expressions obtained from the FA,
\be
\label{eq:FApoisson}
	k^2 \psi_{\textrm{FA}} = - \frac{1}{2 M_{\textrm{pl}}^{2}} a^2 (\rho_{d} \delta_{d, \textrm{FA}} + \rho_{\nu, \textrm{New}} \delta_{\nu, \textrm{FA}})\,.
\ee
The system of Eqs.~(\ref{eq:nuFA}), (\ref{eq:dFA}) and (\ref{eq:FApoisson}) is the so-called {\it two fluid approximation}. What we have then is a first principles derivation of the two-fluid approximation, starting from the Boltzmann equation, and assuming the Newtonian evolution, given by Eq.~(\ref{eq:NewEOM}), of individual dust and neutrino particles. 

\subsection{Non-relativistic fluid (NRF) equations for neutrinos and dust }
\label{subsec:NRF}
To study the impact of treating fast-moving neutrinos as non-relativistic particles, we will go beyond the two-fluid approximation in two ways. First, we will include the anisotropic stress, which should physically be present for a neutrino fluid modeled by N-body particles and allows us to obtain a non-relativistic analog of the CLASS fluid approximation \cite{1104.2935}. Including anisotropic stress also enables us to check that neutrino shear stress has a negligible impact on matter perturbations at late times, which validates the use of the FA to generate initial conditions for the simulations. Second, since our concern here is on SR effects, we will keep all GR terms that are missing in the two-fluid approximation of Sec. \ref{subsec:FA}. This second choice allows the expressions in this section to correctly reproduce the relativistic dynamics on large scales, as $z \to 0$ (see \ref{sec:numericalresults}). This is in contrast to the FA equations in Sec. \ref{subsec:FA}, which disagree with the exact linear theory expressions, on the large scales, even for dust \cite{1605.05283}. 

We will work with the Newtonian gauge metric,
\begin{align}
\label{eq:newmetric}
	ds^2 &= -(1+2\psi)dt^2 + a^2(t)(1-2\phi)d\vec{x}^2\,, \nonumber \\
	 &= a^2(\tau)\left[-(1+2\psi)d\tau^2 + (1-2\phi)d\vec{x}^2\right]\,.
\end{align}
In the non-relativistic (NR) limit ($(v/c)^2\ll 1$), the particle equation of motion (i.e. the geodesic equation) is, 
\be
\label{eq:NRlimitEOM}
\frac{d\vec{v}}{dt}+\left(H-\dot{\phi}\right)\vec{v}+\frac{1}{a}\nabla\psi  = 0\,.
\ee
where the peculiar velocity is now given by,
\be
\label{eq:GRvelocitymain}
\vec{v} = a(1-\phi -\psi)\frac{d\vec{x}}{dt}\,.
\ee
and includes local inhomogeneities in both time and length intervals. Working with the above equation of motion, along with the Newtonian expressions for the energy density and pressure in Eq.~(\ref{eq:densitynew}) and Eq.~(\ref{eq:pressurenew}), produces the following fluid approximation for non-relativistic neutrinos,
\begin{subequations}
\label{eq:nuFA-S}
\begin{align}
\delta'_{\nu, \textrm{NRF}} &= -\theta_{\nu, \textrm{NRF}} + 3 \phi_{\textrm{NRF}}' \,,\\ 
\theta'_{\nu, \textrm{NRF}} &= -\mathcal{H} \theta_{\nu, \textrm{NRF}} + c_{g, \textrm{New}}^{2} k^2 \delta_{\nu, \textrm{NRF}} -k^2 \sigma_{\nu, \textrm{NRF}}  \nonumber \\ &  + k^2 \psi_{\textrm{NRF}}\,,\\ 
\sigma'_{\nu, \textrm{NRF}} &= -\Big(2\mathcal{H}+\frac{3}{\tau}\Big) \sigma_{\nu, \textrm{NRF}} + 8\frac{w_{\textrm{New}}}{1+w_{\textrm{New}}} c_{g, \textrm{New}}^{2} \theta_{\nu, \textrm{NRF}}\,,
\end{align}
\end{subequations}
where $w_{\textrm{New}}$ and $c_{g, \textrm{New}}$ are the equation of state and adiabatic sound speed, as computed using the Newtonian expressions for pressure and energy density in Eqs.~(\ref{eq:densitynew}) and (\ref{eq:pressurenew}). 

The dust component evolves according to
\begin{subequations}
\label{eq:dFA-S}
\begin{align}
\delta'_{d, \textrm{NRF}} &= -\theta_{d, \textrm{NRF}} + 3\phi_{\textrm{NRF}}' \,,\\ 
\theta'_{d, \textrm{NRF}} &= -\mathcal{H} \theta_{d, \textrm{NRF}}  + k^2 \psi_{\textrm{NRF}}\,.
\end{align}
\end{subequations}
And finally, the gravitational potentials are subject to, 
\begin{subequations}
\begin{align}
\label{eq:poissonFA-S}
	& k^2 \phi_{\textrm{NRF}} + 3\mathcal{H} (\phi_{\textrm{NRF}}' + \mathcal{H} \psi_{\textrm{NRF}}) =\\ & - \frac{1}{2 M_{\textrm{pl}}^{2}} a^2 (\rho_{d} \delta_{d, \textrm{NRF}} + \rho_{\nu, \textrm{New}} \delta_{\nu, \textrm{NRF}})\,,\nonumber \\ & k^2(\phi_{\textrm{NRF}}-\psi_{\textrm{NRF}}) = \frac{3}{2M_{\textrm{pl}}^2}a^2 \rho_{\nu, \textrm{New}}\sigma_{\nu, \textrm{NRF}}\,. \label{eq:phiminuspsi}
\end{align}
\end{subequations}

For $z \lesssim 100$, it suffices to apply the approximation $ \phi_{\textrm{NRF}}  \approx \psi_{\textrm{NRF}}$, instead of using Eq.~(\ref{eq:phiminuspsi}) \footnote{Setting $\phi = \psi$ at late times is a simplifying approximation that is implictly made in N-body codes. Indeed, we can argue that the relative difference between $\phi$ and $\psi$ is negligible: On the small scales, we can combine Eqs. (\ref{eq:poissonFA-S}) and (\ref{eq:phiminuspsi}) to arrive at $1- \psi/\phi \propto (\rho_{\nu}/\rho_{\textrm{total}})(\sigma_{\nu}/\delta_{\textrm{total}})$, i.e. the relative difference is a product of two small quantities, and is vanishingly small. For $M_{\nu} = 0.15$eV, we found that $|1-\psi_{\textrm{CLASS}}/\phi_{\textrm{CLASS}}| \lesssim \mathcal{O}(10^{-4})$ for $k \gtrsim 0.01$ Mpc$^{-1}$, for all redshift. On the large scales, the neutrino anisotropic stress is vanishing at low redshift, and indeed we find that the error associated to having $\phi \neq \psi$ is subleading in comparison to the systematic error associated to neglecting radiation perturbations at $z \sim 100$. This argument can be applied in both cases of non-relativistic and relativistic neutrinos.} Let us call Eqs.~(\ref{eq:nuFA-S}), (\ref{eq:dFA-S}) and (\ref{eq:poissonFA-S}) the non-relativistic fluid (NRF) equations. To summarize, the NRF equations describe co-evolution of dust and neutrinos evolving according to the non-relativistic limit ($(v/c)^2 \ll 1$) of both the geodesic and Einstein equations. 

\subsection{Relativistic fluid equations (RFE) for neutrinos and dust }
\label{subsec:RF}
The equations in Sec. \ref{subsec:NRF} can be compared with the fully-relativistic CLASS non-cold dark matter (NCDM) fluid approximation \cite{1104.2935}. These expressions continue to use the Newtonian gauge metric, Eq.~(\ref{eq:newmetric}), but keep all terms $\mathcal{O}((v/c)^2)$. In this case the particle equation of motion is, 
\be
\frac{1}{\gamma} \frac{d}{dt}\left(\gamma \vec{v}\right) +\left(H-\dot{\phi}\right)\vec{v}+\frac{1}{a}\nabla\psi +\frac{1}{a}\vec{v}\times\left(\nabla \phi \times \vec{v}\right) =0\,,
\ee
where $v$ continues to be defined through Eq.~(\ref{eq:GRvelocitymain}). The neutrino fluid equations are,
\begin{subequations}
\label{eq:nuCLASS}
\begin{align}
\delta'_{\nu, \textrm{RFE}} &=-(1+w_{\textrm{rel}})(\theta_{\nu,\textrm{RFE}} - 3 \phi'_{\textrm{RFE}}) \\ & \nonumber -3\mathcal{H}(c_{g,\textrm{rel}}^2 - w_{\textrm{rel}})\delta_{\nu,\textrm{RFE}}\,, \\ 
\theta'_{\nu, \textrm{RFE}} &= -\Big[\mathcal{H}(1-3w_{\textrm{rel}}) + \frac{w'_{\textrm{rel}}}{1+w_{\textrm{rel}}}\Big]\theta_{\nu, \textrm{RFE}} + \\ & \frac{c_{g, \textrm{rel}}^{2}}{1+w_{\textrm{rel}}} k^2 \delta_{\nu, \textrm{RFE}}  -k^2 \sigma_{\nu, \textrm{RFE}} +k^2 \psi_{\textrm{RFE}} \,,\nonumber \\
 \sigma'_{\nu, \textrm{RFE}} &= -\Big\{ \Big[ (2-3w_{\textrm{rel}})-\frac{\mathcal{P_{\textrm{rel}}}}{P_{\textrm{rel}}}\Big]\mathcal{H} + \frac{w'_{\textrm{rel}}}{1+w_{\textrm{rel}}} + \frac{3}{\tau}\Big\} \sigma_{\nu, \textrm{RFE}}  \\ + & 8\frac{w_{\textrm{rel}}}{1+w_{\textrm{rel}}} c_{g, \textrm{rel}}^{2} \theta_{\nu, \textrm{RFE}} \,,\nonumber
\end{align}
\end{subequations}
where $w_{\textrm{rel}}=P_{\textrm{rel}}/\rho_{\textrm{rel}}$, 
\be
\label{eq:hmwp}
	\mathcal{P_{\textrm{rel}}} = \frac{4\pi}{3} ma^{-3} \int_{0}^{\infty} q^2f_{0}(q) dq  \Big(\frac{q}{ma}\Big)^4 \frac{1}{\big[1+\big(\frac{q}{ma}\big)^2\big]^\frac{3}{2}}\,,
\ee
is a higher velocity weight background pressure, and
\be
	c_{g, \textrm{rel}}^2 = \frac{5}{3} \frac{w_{\textrm{rel}}}{1+w_{\textrm{rel}}} \Big(1-\frac{1}{5}\frac{\mathcal{P_{\textrm{rel}}}}{P_{\textrm{rel}}}\Big)\,,
\ee
is the relativistic expression for the adiabatic sound speed squared, as follows from Eq.~(\ref{eq:cg}), Eq.~(\ref{eq:densityrel}), Eq.~(\ref{eq:pressurerel}), and Eq.~(\ref{eq:hmwp}). We will refer to this as the relativistic fluid equations (RFE). 

The dust RFE are identical to those in Eq.~(\ref{eq:dFA-S}), with the potentials replaced by the ones sourced by the full (kinetic plus rest mass) energy density of neutrinos, 
\begin{subequations}
\label{eq:drel}
\begin{align}
\delta'_{d,\textrm{RFE}} &= -\theta_{d,\textrm{RFE}} + 3\phi'_{\textrm{RFE}} \,,\\ 
\theta'_{d,\textrm{RFE}} &= -\mathcal{H} \theta_{d,\textrm{RFE}}  + k^2 \psi_{\textrm{RFE}}\,.
\end{align}
\end{subequations}
where $\phi_{\textrm{RFE}}$, $\psi_{\textrm{RFE}}$ are subject to, 
\begin{subequations}
\begin{align}
\label{eq:poissonRFE}
	& k^2 \phi_{\textrm{RFE}} + 3\mathcal{H} (\phi'_{\textrm{RFE}} + \mathcal{H} \psi_{\textrm{RFE}}) =  \nonumber \\ & - \frac{1}{2 M_{\textrm{pl}}^{2}} a^2 (\rho_{d} \delta_{d,\textrm{RFE}} + \rho_{\nu, \textrm{rel}} \delta_{\nu, \textrm{RFE}})\,, \\
	  & k^2(\phi_{\textrm{RFE}}-\psi_{\textrm{RFE}}) = \frac{3}{2M_{\textrm{pl}}^2}a^2 (\rho_{\nu, \textrm{rel}}+P_{\nu, \textrm{rel}}) \sigma_{\nu, \textrm{RFE}} \,.\label{eq:phiminuspsirel}
\end{align}
\end{subequations}
That is, including the relativistic kinetic energy of neutrinos as a source to the gravitational potential. Also, and again for $z \lesssim 100$, it suffices to set $\phi_{\textrm{RFE}} \approx \psi_{\textrm{RFE}}$, instead of using Eq.~(\ref{eq:phiminuspsirel}).

\section{Numerical Results}
\label{sec:numericalresults}
The overall difference between the output of Boltzmann codes, and the two-fluid approximation presented in Sec. \ref{subsec:FA}, is described in \cite{1605.05283}. Here we study the errors induced specifically from neglecting special relativistic corrections, by comparing our non-relativistic fluid equations, of Sec. \ref{subsec:NRF}, to the relativistic fluid equations described in Sec. \ref{subsec:RF}. As discussed in Sec. \ref{sec:NewtonianneutrinosBG}, we expect errors in the NRF associated to the overestimation in the neutrino free-streaming scale $\lambda_{\textrm{fs}}$, and corresponding to an overestimation in the suppression to the growth of structure, caused by particles moving too fast in the simulation box, as seen in Figures \ref{fig:velocitydist005}-\ref{fig:cs}. 

We consider the specific example of a flat $\nu \Lambda$CDM universe, with three neutrino components in the degenerate case of equals masses, with total mass $M_{\nu}$. Our choice of cosmological parameters is defined by: $h=0.67$, $\omega_{\textrm{d}}=\Omega_{\textrm{d},0}h^2=0.1424$, $T_{\textrm{\textrm{CMB}},0} = 2.725 K$, $T_{\nu,0} = 1.95K$. The neutrino masses are left as free parameters and $\Omega_{\Lambda,0}$ is determined from the constraint equation $\sum_{I} \Omega_{I,0} = 1$. 

We will be interested in a redshift range of $z \lesssim 100$, relevant for the study of structure formation. In this case, both CDM and baryons can be treated as a single dust component, and radiation may be ignored. This last assumption leads to the introduction of significant systematic errors in the computation of observables for larger values of redshift, and on the largest scales. In order to avoid this, along with other sources of systematic error on the large scales, initial conditions are often set in N-body simulations from the rescaling of the matter power spectrum at $z=0$, to the simulation initial redshift $z_{i}$. This procedure enforces recovery of the predictions of the linear theory on linear scales, and low redshift, at the expense of predictivity at high redshift \cite{1605.05283}. However, it has been argued that this rescaling is not feasible in some cases \cite{1702.03221}, including the example of cosmologies with massive neutrinos, though one is still able to choose the initial conditions in a suitable way \cite{1810.12019}. Taking this into account, and since our goal is to compare non-relativistic with relativistic neutrinos, and not necessarily reproduce the output of a Boltzmann code (so we don't need to worry about the contribution from radiation or lack thereof), we here choose the initial conditions for the transfer functions at $z_{i}=100$ or $z_{i}=60$. Notwithstanding, we will comment on how a rescaling procedure (or other suitable methods to generate the initial conditions) would impact our results. 

The initial conditions for dust and gravitational potential $\phi$ are generated in the same way for both the RFE and NRF systems, i.e. directly from the output of CLASS, at the initial redshift (the dust transfer functions, at the initial time, are set to be a weighted average of CDM and baryon transfer functions). We do the same for neutrinos in the RFE, that is
\begin{subequations}
\label{eq:ICRFE}
\begin{align}
	& \delta_{\nu,\textrm{RFE}}(z_{i}) = \delta_{\nu,\textrm{CLASS}}(z_{i})\,, \\
	& \theta_{\nu,\textrm{RFE}}(z_{i}) = \theta_{\nu,\textrm{CLASS}}(z_{i})\,, \\
	& \sigma_{\nu,\textrm{RFE}}(z_{i}) = \sigma_{\nu,\textrm{CLASS}}(z_{i})\,.
\end{align}
\end{subequations}
	 
For neutrinos in the NRF, however, we have to be more careful. This is because the non-relativistic fluid equations are solving for a different set of variables, e.g. the neutrino mass density as opposed to the energy density. For adiabatic initial conditions, the large-scale super-horizon initial conditions satisfy $\delta_{\nu,\textrm{RFE}} \approx (1+w_{\textrm{rel}})\delta_{d,\textrm{RFE}}$. On the other hand, the correct evolution of the mass density on large scales would be given by $\delta_{\nu,\textrm{NRF}} \approx \delta_{d,\textrm{NRF}}$. On sub-horizon scales the neutrino perturbations are no longer adiabatic. For the examples in this paper, however, we find that the solutions for neutrino perturbations on sub-horizon scales quickly lose sensitivity to the initial conditions once the neutrinos become non-relativistic \cite{1003.0942}. The initial conditions for the neutrino fluid then have a negligible impact on the evolution of matter perturbations on small scales.

This allows us to choose the initial conditions for neutrinos in the NRF system as follows,
\begin{subequations}
\label{eq:ICNRF}
\begin{align}
	& \delta_{\nu,\textrm{NRF}}(z_{i}) = \frac{\delta_{\nu,\textrm{CLASS}}(z_{i})}{1+w_{\textrm{rel}}(z_{i})}\,, \\ 
	& \theta_{\nu,\textrm{NRF}}(z_{i}) = \frac{\theta_{\nu,\textrm{CLASS}}(z_{i})}{1+w_{\textrm{rel}}(z_{i})}\,, \\ 
	& \sigma_{\nu,\textrm{NRF}}(z_{i}) = \frac{\sigma_{\nu,\textrm{CLASS}}(z_{i})}{1+w_{\textrm{rel}}(z_{i})}\,,
\end{align}
\end{subequations}
where we apply the same procedure for $\theta_{\nu,\textrm{NRF}}$ and $\sigma_{\nu,\textrm{NRF}}$ in order to correct for the fact that their definition differs from the relativistic expression by a factor of $(1+w_{\textrm{rel}})$ (see the appendix \ref{sec:fluidderiv}). 

\begin{figure*}
		\centering
		\includegraphics[width=1\textwidth]{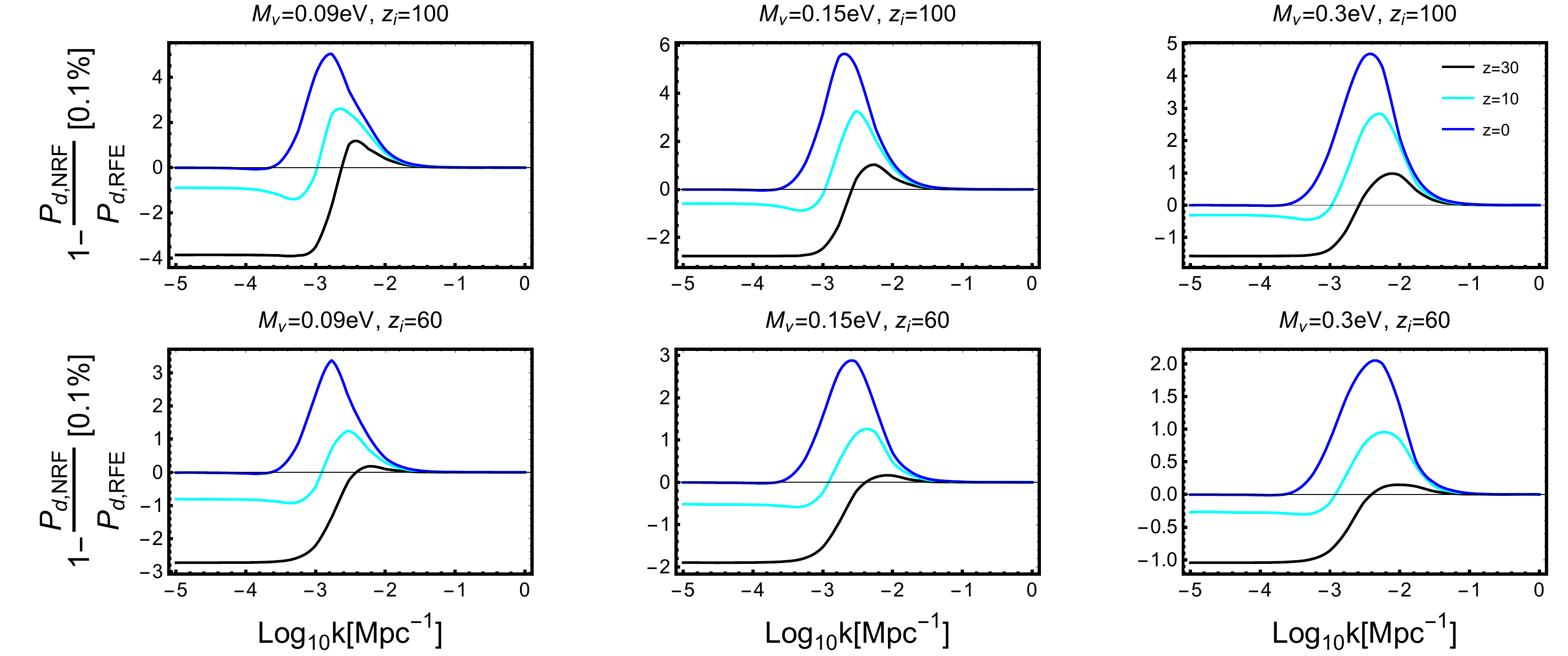}
		\caption{Relative difference in the dust (CDM+b) power spectrum, $P = |\delta_{d}|^2$, between the NRF and RFE, corresponding to Newtonian and relativistic neutrinos, respectively. The six plots are obtained combining two possible values for the initial redshift, $z_{i}=60$ or $z_{i}=100$, along with three possible choices for the total neutrino mass $M_{\nu} =0.09$eV, $M_{\nu} = 0.15$eV or $M_{\nu} =0.30$eV, ranging from (approximately) the smallest to the largest possible value, according to constraints from cosmology and neutrino oscillations. The shift in the neutrino free-streaming scale induces a $\lesssim$0.5\% error in the linear matter power spectrum, in all cases, at around the neutrino horizon scale. } 
\label{fig:ps}
\end{figure*}

We are now ready to state our results. In Fig. \ref{fig:ps} we compare the dust power spectra in universes where SR terms for neutrinos are included or ignored, computed using the RFE (Sec. \ref{subsec:RF}) and NRF (Sec. \ref{subsec:NRF}), respectively. These plots deserve careful explanation. Since NRF neutrinos move faster, they wash-out power on larger scales than neutrinos evolved with the RFE, and hence we expect to see an overestimation in the suppression to the growth of structure. In terms of our curves, this implies that they should be above zero up to the horizon of Newtonian neutrinos, with a peak in between the Newtonian and relativistic horizons. In fact, notice that the peaks in the curves shift only slightly towards lower $k$ as $z \to 0$, following the trend expected from the neutrino particle horizon, since its integral is dominated by early times when $c_{g,\textrm{New}} $ is most different from $c_{g,\textrm{rel}}$. Furthermore, the position of the peak shifts towards smaller scales as we increase the neutrino mass, as expected. Also notice that, at the initial redshift, the differences are zero by construction, while at later times the free-streaming scales of Newtonian and relativistic neutrinos approach one another. That explains why we see no feature on scales where both NRF and RFE neutrino perturbations get completely washed-out, and the neutrino component has no effect on dust. 

The plots in Fig.  \ref{fig:ps} also show significant disagreement on large scales and for higher values of redshift. This is because the RFE and NRF system are solving for different variables, i.e. neutrino energy density including or ignoring the kinetic energy, respectively. On the large scales, this introduces a $ \propto (1+w_{\textrm{rel}})$ disagreement on the neutrino density contrast transfer function, as previously discussed, which impacts the evolution of dust through gravitational coupling. This difference disappears as  $w_{\textrm{rel}}\rightarrow 0$ at late times.

Another interesting feature observed in Fig. \ref{fig:ps} is how the error in the matter power spectrum depends on the neutrino mass scale and the initial redshift of the simulation (for our exercise, this amounts to the redshift at which we start solving the NRF). In agreement with intuition, the error is larger for earlier starting redshifts $z_i$, since the temperature is larger in comparison to the mass at the initial redshift, and hence the NR approximation becomes worse. On the other hand, for $z_{i}=100$, varying the mass seems to leave the size of the peak (approximately) unaltered. We attribute this to two competing effects: for smaller masses the NR approximation is worse, but the overall effects of neutrinos on the power spectrum are also smaller. These two effects don't always cancel each other out, as one can see from the plots with the initial redshift of $z_{i}=60$. 

The errors in the linear power spectrum illustrated in Fig  \ref{fig:ps} are $\lesssim 0.5\%$ for the masses and initial redshifts values we consider, and are associated to an overestimation in the neutrino horizon. The effects of neutrino masses in cosmology are traditionally illustrated in the ratio of matter power spectra between cosmologies with and without neutrino masses. For scales above the neutrino horizon, $k<k_{\textrm{h}}$, it assumes the value $1$, while on small scales for which $k \gg k_{\textrm{h}}$, below the free-streaming scale, it can be approximated as follows,
\be
\label{eq:nsp}
	\frac{P_{d}(M_{\nu})}{P_{d}(M_{\nu}=0)}  \approx 1-6f_{\nu}, \ \ k \gg k_{\textrm{h}}
\ee
where $\Omega_{\textrm{m}} = \Omega_{\textrm{d}} + \Omega_{\nu}$ is kept fixed, and $f_{\nu} = \Omega_{\nu}/\Omega_{\textrm{m}}$. This is the neutrino-induced suppression to the growth of structure, and it is actually enhanced by non-linear effects \cite{0802.3700}. Notice that massless neutrinos behave like photons, with their contribution to the energy budget of the universe becoming negligible at late times. In this case it suffices to simply concentrate all of the matter in the dust component, i.e. to set $\Omega_{\textrm{m}} = \Omega_{\textrm{d}}$ in the model with massless neutrinos.

\begin{figure}
		\centering
		\includegraphics[width=0.45\textwidth]{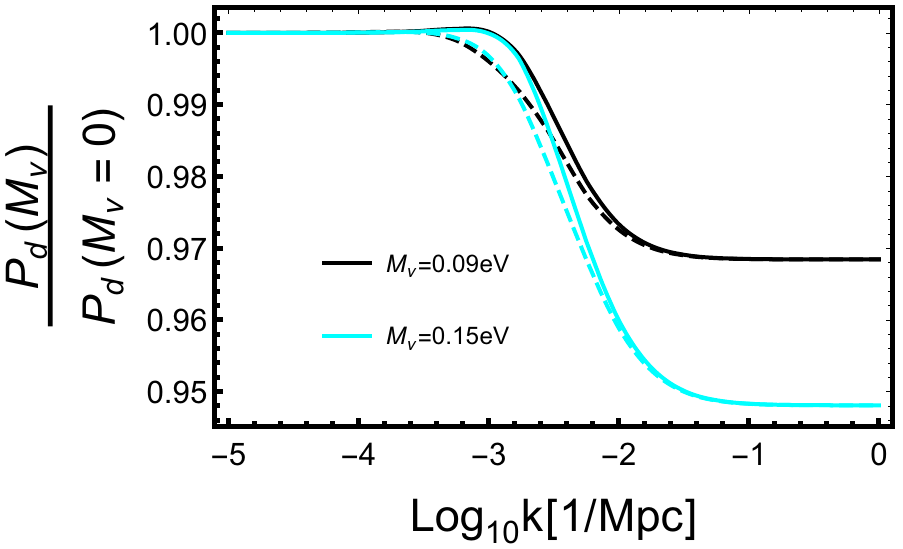}
		\caption{Ratio of matter (cdm+b) power spectra between models with massive and massless neutrinos, at $z=0$, for the neutrino mass scales of $M_{\nu}=0.09$eV and $M_{\nu}=0.15$eV. The solid lines corresponds to massive neutrinos evolved with the RFE, while dashed lines are computed with the NRF. This also illustrates the overestimation in the neutrino sound horizon, while emphasizing that the neutrino-induced suppression to the growth of structure, in the small scales, is correctly reproduced. The overestimation in the neutrino horizon causes, however, an up to $\sim 10\%$ correction to the shape of the neutrino-induced suppression of the cold dark matter power spectrum, with the largest errors occurring for the lowest values of $M_\nu$.} 
\label{fig:rps}
\end{figure}

The ratio of power spectra between cosmologies with and without neutrino masses at $z=0$, and simulation initial redshift of $z_{i} = 100$, is plotted in Fig. \ref{fig:rps} for the neutrino mass scales of $M_{\nu}=0.09$eV and $M_{\nu}=0.15$eV. The model with massive neutrinos is computed with both the RFE and NRF systems, while the model with massless neutrinos is computed using only the RFE. As can be seen in Fig. \ref{fig:rps}, the neutrino horizon is overestimated when using non-relativistic equations of motion, in agreement with Fig. \ref{fig:ps}. For the smallest neutrino mass we consider, $M_\nu = 0.09$eV, this causes an up to $\sim 10\%$ correction to the shape of the neutrino-induced suppression to the cold dark matter power spectrum. For heavier masses the correction to the shape is smaller. We see no shift in the location of the transition to the constant suppression of the linear power spectrum at high $k$. This is because this feature is set by the free-streaming scale, and its overestimation disappears in the limit $z \to 0$.

\begin{figure}
		\centering
		\includegraphics[width=0.45\textwidth]{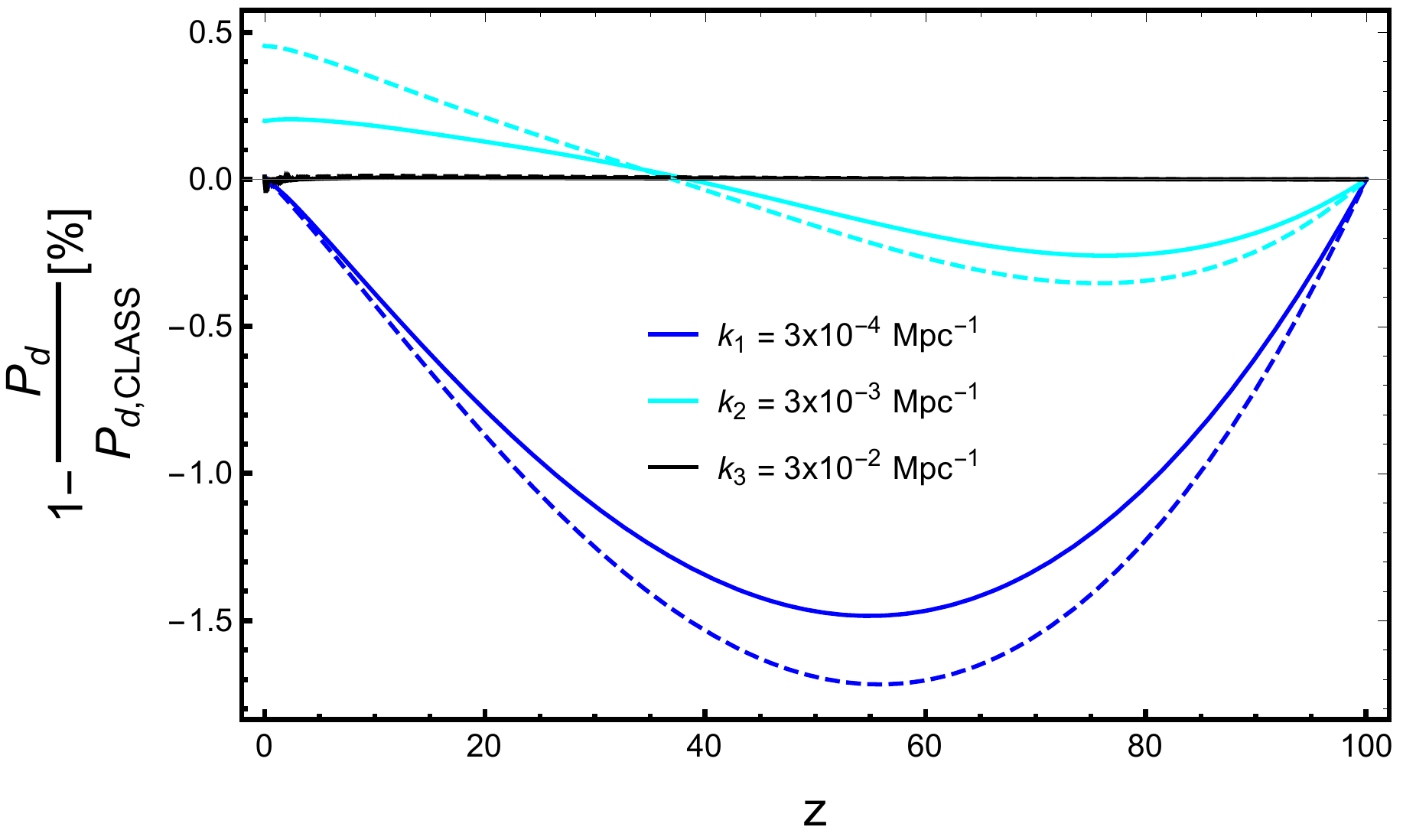}
		\caption{Relative difference between dust (cdm+b) power spectra, computed with the RFE (solid lines) or NRF (dashed lines), and CLASS. We consider three fixed values of $k$: $k_{1} = 3 \times 10^{-4} \textrm{Mpc}^{-1} < k_{\textrm{h}}$ (blue), above the neutrino horizon, $k_{2} = 3 \times 10^{-3} \textrm{Mpc}^{-1} \approx k_{\textrm{h}}$ (cyan), and $k_{3} = 3 \times 10^{-2} \textrm{Mpc}^{-1} > k_{\textrm{h}}$ (black). The neutrino mass scale is fixed at $M_{\nu} = 0.15$eV, the simulation initial redshift at $z_{i}=100$, and the relative differences are plotted as functions of redshift. For $k > k_{\textrm{h}}$, both the RFE and NRF exactly reproduce CLASS, while for $k \lesssim k_{\textrm{h}}$, the difference between RFE and CLASS is dominated by the contribution from radiation, which we have neglected.} 
\label{fig:vsCLASS}
\end{figure}

Finally, a direct comparison of the NRF and RFE with CLASS is  illuminating, for it shows the impact of other sources of systematic error in N-body simulations, that are neglected in our analysis. This is presented in Fig. \ref{fig:vsCLASS}, for a neutrino mass scale of $M_{\nu} = 0.15$eV and simulation initial redshift $z_{i}=100$. 

For $k_{1} = 3 \times 10^{-4} \textrm{Mpc}^{-1} < k_{\textrm{h}}$, the difference between the RFE and CLASS is dominated by the contribution from radiation (with a subleading effect coming from the fact that $\phi \neq \psi$), while the additional difference between the RFE and NRF is associated to having different definitions of neutrino energy densities, i.e. including or not the kinetic energy. For $k_{2} = 3 \times 10^{-3} \textrm{Mpc}^{-1} \approx k_{\textrm{h}}$, the difference between the RFE and CLASS is again dominated by the contribution from radiation, while the difference between the RFE and NRF is now due to the differences in definitions of fluid properties, as before (this dominates at $z \gg 1$), but also from the overestimation of the neutrino horizon (this dominates at $z \to 0$). This explains the intersection of NRF and RFE curves as the transition between these two regimes. For $k_{3} = 3 \times 10^{-2} \textrm{Mpc}^{-1} > k_{\textrm{h}}$, both the NRF and RFE reproduce CLASS exactly, in agreement with our previous results. The differences between the NRF and RFE with CLASS, found to be significant for $k \lesssim k_{\textrm{h}}$ (at linear scales), can be alleviated for all redshifts when interpreting the output of the simulations in a suitable gauge \cite{1807.03701}. An alternative procedure would be to rescale the power spectrum from $z=0$ to the simulation initial redshift $z_{i}$ \cite{1605.05283}, to generate initial conditions for the simulations. This alleviates the errors at $z \to 0$, but increase the errors at high redshift $z \gg 1$.

\section{Implications for N-body Simulations with Massive Neutrinos}
\label{sec:implications}

The systematic errors associated to the overestimation of the free-streaming scale, at higher redshifts, are only present at around the scale of neutrino horizon as $z \to 0$, safely within the regime of applicability of the linear theory, as can be seen in Figs. \ref{fig:ps} and \ref{fig:rps}. This is in agreement with the results of \cite{2003.07387}, where the authors add the effects of neutrinos in linear theory through a post-processing after running the simulations. There is, however, a procedure that enforces that the outputs of N-body simulations will match the linear theory results on linear scales (with non-linear corrections on top), i.e. the rescaling of the power spectrum, from $z=0$ to $z_{i}$, to generate initial conditions for the simulations \cite{1605.05283, 1810.12019} (see the discussion in Sec. \ref{sec:numericalresults}). Note that this procedure can only force agreement with linear theory at single redshift, typically chosen to be $z=0$. An alternative scenario is to interpret the outputs of the simulation in a suitable gauge, which allows N-body simulations to approximate the dynamics of both dust and neutrinos to high accuracy \cite{1807.03701}.

There is still the question of to what extent the non-linear corrections, generated by N-body simulations, are accurate. This is because the changes to the neutrino velocity distribution, shown in Figs. \ref{fig:velocitydist005} and \ref{fig:velocitydist02}, can also potentially cause errors in nonlinear clustering of neutrinos. However, since the relativistic and Newtonian velocity distributions match in the low-velocity end, we expect these errors to be negligible. While this is hard to compute analytically, one can get a simple estimate of the neutrinos that end up nonlinearly clustered (i.e. bound) in dark matter halos as follows. Consider a halo of mass $M$.  Only neutrinos with $v < v_{\textrm{\textrm{esc}}}$, where $v_{\textrm{\textrm{esc}}}$ is the escape velocity of the halo, will end up bound. It then follows that out of all the neutrinos that encounter the halo, only a fraction
\be
\label{eq:fnb}
f_{\textrm{bound}}(v_{\textrm{esc}}) \propto \int_0^{p_{\textrm{esc}}} dp \ p^2 f_{0}(p)
\ee 
will end up bound to the halo (see, e.g. \cite{1310.6459}). For a halo of mass $M$ (comprised of dust), the total neutrino mass in the vicinity of the halo is $M_\nu = (4/3)\pi R_L^3 \rho_{\nu,\textrm{\textrm{New}}}$, where $R_L = (M/(4/3\pi \rho_{\textrm{d}}))^{1/3}$ is the Lagrangian radius of the halo. An estimate of the neutrino mass bound to a halo of mass $M$ is then
\begin{align}
M_{\nu, \textrm{bound}} &= f_{\textrm{bound}}(v_{\textrm{esc}}) \frac{4}{3}\pi R_L^3 \rho_{\nu,\textrm{New}}\,\\
& = f_{\textrm{bound}}(v_{\textrm{esc}})  \frac{\rho_{\nu,\textrm{New}}}{\rho_{\textrm{d}}} M
\end{align}

\begin{figure*}
		\centering
		\includegraphics[width=0.9\textwidth]{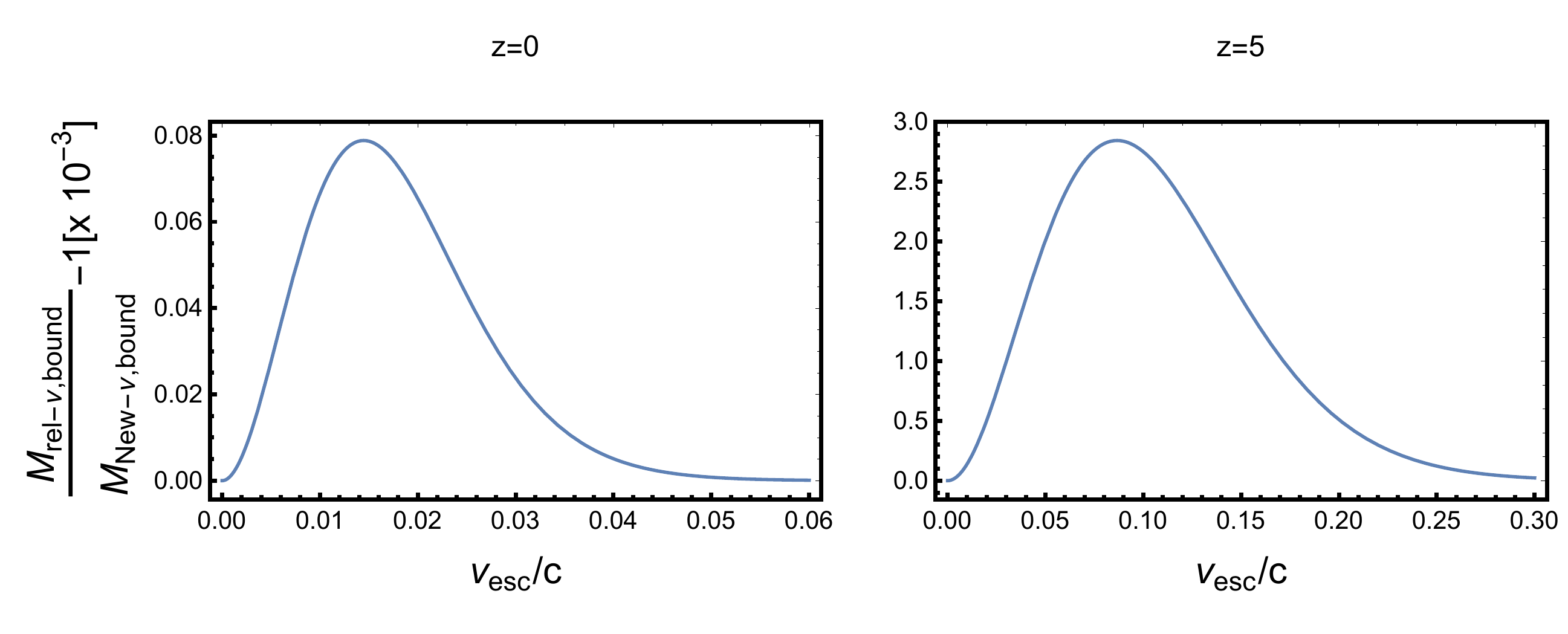}
		\caption{Relative difference between relativistic and Newtonian neutrino masses bound to a dark matter halo, as follows from the relativistic or Newtonian velocity distribution functions, for $m_{\nu} = 0.05$eV. Particle-based Newtonian N-body simulations accurately model non-linear clustering effects of neutrinos. } 
\label{fig:ratiofnb}
\end{figure*}

The escape velocity can be estimated by $v_{\textrm{esc}}^2 \sim GM/R_L \sim M/M_{\textrm{pl}}^2 R_L$. The ratio of relativistic and Newtonian neutrino masses bound to a halo then reads 
\be
\label{eq:ratiofnb}
\frac{M_{\textrm{rel}-\nu, \textrm{bound}}}{M_{\textrm{New}-\nu, \textrm{bound}}} = \frac{f_{\textrm{rel,bound}}(v_{\textrm{esc}})}{f_{\textrm{New,bound}}(v_{\textrm{esc}})}
\ee
where $p_{\textrm{esc}} = \gamma_{\textrm{esc}} m_{\nu} v_{\textrm{esc}}$ or $p_{\textrm{esc}} = m_{\nu} v_{\textrm{esc}}$ in Eq.~(\ref{eq:fnb}) for relativistic and Newtonian neutrinos, respectively. The relative difference in Eq.~(\ref{eq:ratiofnb}) is plotted as a function of the escape velocity in Fig. \ref{fig:ratiofnb}, for redshifts of $z=0$ and $z=5$, and a neutrino mass of $m_{\nu} = 0.05$eV. As expected, the error in non-linear clustering of neutrinos is negligible. Therefore, particle-based N-body simulations accurately model the physics of non-linear structure formation in the presence of massive neutrinos, while the overestimation in the neutrino horizon, along with other soucers of systematic error on linear scales, can be alleviated by either using the rescaling procedure to generate initial conditions for the simulations, or interpreting the outcome of the simulations in a suitable gauge. For smaller neutrino masses, the sound horizon is pushed to even larger scales where the linear theory produces more accurate results, while for larger masses the non-relativistic approximation becomes better, and the error in the power spectrum smaller. Hence, our results hold for any simulation initial redshift of $z_{i} \lesssim 100$, and degenerate neutrino masses in the allowed range from neutrino oscillation experiments and constraints from cosmology. For the lightest neutrino mass states with nonzero mass ($m_\nu \approx 0.01$ eV for the minimal mass normal ordering), the errors due to neglecting SR effects on the neutrino evolution are more severe, but these light neutrinos would also comprise a smaller fraction of the neutrino energy density and so the overall errors on the dust perturbations should be smaller.

\section{Conclusions}	
\label{sec:conclusions}

We have argued that particles with large thermal velocities, such as neutrinos, evolving with the Newtonian equation of motion travel faster and further than they would if evolved with the correct, relativistic equation of motion. This causes an overestimation in the neutrino free-streaming scale, and could potentially introduce errors in both linear and nonlinear clustering of neutrinos, in particle-based N-body simulations of structure formation. This is a special relativistic effect of neutrinos, that is neglected in N-body simulations, and has not been systematically studied. 

In order to determine the impact of this on the evolution of matter perturbations, we developed the exact linear-theory evolution of inhomogeneities in the distribution of both Newtonian and non-relativistic neutrinos, where the non-relativistic equation of motion adds large scale general relativistic corrections to the Newtonian equation of motion. We apply our findings to derive the usual two-fluid approximation, that is used to generate initial conditions for N-body simulations by rescaling of the matter power spectrum from $z=0$ to the initial simulation redshift $z_i$ \cite{1605.05283}, along with a fluid approximation for non-relativistic neutrinos that is analogous to its fully-relativistic counterpart used in the code CLASS \cite{1104.2935}.

It was found that the overestimation in the neutrino free-streaming scale has a sub-percent impact on the linear matter (i.e. CDM +baryon) power spectrum, for neutrino mass scales in the allowed range from neutrino oscillation experiments and constraints from cosmology, and for an initial simulation redshift of $z_{i} \lesssim 100$. This error is at around the scale of neutrino horizon, safely within the regime of applicability of the linear theory. On the other hand, the free-streaming scale approaches the non-linear regime for higher masses and smaller redshifts, in which case the non-relativistic dynamics becomes a good approximation, and the shift in the free-streaming scale disappears in the limit $z \to 0$. As a consequence, N-body simulations accurately model non-linear clustering of neutrinos. Approaches to limit errors on the linear scales are to rescale the power spectrum from $z=0$ to the simulation initial redshift $z_{i}$ \cite{1605.05283, 1810.12019}, to generate initial conditions for the simulations that give the correct $z=0$ power spectrum, while having some errors at $z \gg 1$. Alternatively, one may also interpret the output of the simulations in a suitable gauge that allows the simulations to approximate the evolution of both neutrinos and CDM \cite{1807.03701}.

Particle-based implementations of neutrinos in N-body simulations are then powerful tools to accurately model the non-linear formation of structure in our universe. Including neutrinos as N-body particles has the advantage of accounting for neutrino non-linear clustering effects, with the shortcoming of having to deal with the well-known problem of shot noise, due to large thermal velocities (however, there are ways around it, e.g. \cite{1801.03906, 1807.03701}). The other possibility, to include neutrinos as a linear component, completely misses effects of non-linear clustering of neutrinos \cite{1811.00904}. However, such effects might be negligible for the small neutrino masses constrained by cosmology, though it has been argued that the slow tail in the distribution of neutrinos cluster strongly enough to warrant a non-linear treatment \cite{2011.12503}.

\acknowledgements
We are grateful to Yin Li and Francisco Villaescusa-Navarro for helpful correspondence. This work is supported by DOE DE-SC0017848.	
	
\appendix

\section{Anisotropies of Newtonian and non-relativistic NCDM components}
\label{sec:fluidderiv}

We are interested in scalar perturbations of FRW, working in the Newtonian gauge. The metric reads
\begin{align}
	ds^2 &= a^2(\tau)[-(1+2\psi)d\tau^2 + (1-2\phi)d\vec{x}^2] \\
	&= -(1+2\psi)dt^2 + a^2(t)(1-2\phi)d\vec{x}^2
\end{align}
The geodesic equation gives the equation of motion for a point particle as, 
\be
\frac{1}{\gamma} \frac{d}{dt}\left(\gamma \vec{v}\right) +\left(H-\dot{\phi}\right)\vec{v}+\frac{1}{a}\nabla\psi +\frac{1}{a}\vec{v}\times\left(\nabla \phi \times \vec{v}\right) =0
\ee
where the peculiar velocity is given by
\be
\label{eq:GRvelocity}
\vec{v} = a(1-\phi -\psi)\frac{d\vec{x}}{dt}
\ee
and $1/\gamma^2 = 1-v^2$. In the Non-relativistic (NR) limit ($(v/c)^2\ll 1$), this reduces to
\be
\label{eq:geoNRlimit}
\frac{d\vec{v}}{dt}+\left(H-\dot{\phi}\right)\vec{v}+\frac{1}{a}\nabla\psi  = 0\,.
\ee

Let us contrast Eq.~(\ref{eq:geoNRlimit}) with Newton's second law, in the presence of a peculiar gravitational field $\psi$. It reads, in terms of comoving coordinates $\vec{x}(t)$, 
\be
\label{eq:geoNEWlimit}
\frac{d\vec{v}}{dt} + H\vec{v} = -\frac{1}{a} \vec{\nabla} \psi\,.
\ee
where proper (physical) coordinates are given by $\vec{r}(t) = a(t)\vec{x}(t)$, and here we define the peculiar velocity 
\be
\label{eq:Newvelocity}
\vec{v} = a \frac{d\vec{x}}{dt}
\ee
This is the equation of motion used in particle-based N-body simulations. It can be obtained from Eq.~(\ref{eq:geoNRlimit}), after dropping the GR term $\dot{\phi}$, along with the factor of $(1-\phi -\psi)$ in Eq.~(\ref{eq:GRvelocity}), that accounts for inhomogeneities in both position and time intervals. It contributes to Eq.~(\ref{eq:geoNRlimit}) with further derivatives of the potential and order $(v/c)^2$ terms. From this we conclude that Eq.~(\ref{eq:geoNRlimit}) and Eq.~(\ref{eq:geoNEWlimit}) agree on small scales inside the horizon, in the NR limit we are considering. 

We will first consider the linear-theory evolution of a Newtonian NCDM component, according to Eq.~(\ref{eq:geoNEWlimit}) and Eq.~(\ref{eq:Newvelocity}). The equation of motion can be derived from the action
\begin{align}
\label{eq:Newaction}
	S_{\textrm{New}} &= \int dt L_{\textrm{New}}  \nonumber \\ &=  m\int dt \left[\frac{1}{2} a^2 \left(\frac{d\vec{x}}{dt}\right)^2 - \psi\right]
\end{align}

The canonical momentum associated to this is
\be
	\vec{\Pi}_{\textrm{New}} =  \frac{\partial L_{\textrm{New}}}{\partial \dot{\vec{x}}} = ma^2 \frac{d\vec{x}}{dt} = ma\vec{v} = \vec{q}
\ee
where $\vec{q} = a\vec{p}$, $\vec{p}$ is the proper momentum, and spatial indices are raised and lowered with the kronecker delta. We may now find the Hamiltonian associated to Eq.~(\ref{eq:Newaction}), and from it Hamilton's equation of motion. It yields
\be
\label{eq:dynamicsphasespace}
	\frac{d\vec{q}}{d\tau} = -ma\vec{\nabla} \psi\,.
\ee

The dynamics of the distribution function follows from the collisionless Boltzmann equation
\be
	\frac{df}{d\tau} = \frac{\partial f}{\partial \tau} + \frac{dx^{i}}{d\tau} \frac{\partial f}{\partial x^{i}} + \frac{dq^i}{d\tau} \frac{\partial f}{\partial q^{i}} = 0
\ee

Splitting the distribution function as $f=f_{0}(q)(1+\Psi(\tau, \vec{x}, \vec{q}))$, working to leading order on perturbations (e.g. $\Psi$, $\phi$, $\psi$), and moving to momentum space ($\partial_{i} \to ik_{i}$)
\be
\label{eq:Boltzmann}
	\Psi' + i\frac{qk}{ma} (\hat{n} \cdot \hat{k}) \Psi = i\frac{mak}{q} (\hat{n} \cdot \hat{k}) \psi \frac{d\ln f_{0}}{d\ln q}\,,
\ee
where $\hat{n} = \vec{q}/q$ and $'$  denotes partial derivative with respect to conformal time. Next we make the usual assumption of axial symmetry of $\Psi$ around $\hat{k}$, at the initial redshift. This allows one to write the usual multipole expansion
\be
\label{eq:mexp}
\Psi(\tau,\vec{k},q,\hat{n}) = \sum_{l=0}^{\infty} (-i)^{l}(2l+1)\Psi_{l}(\tau,k,q) P_{l}(\hat{n} \cdot \hat{k})
\ee
where $P_\ell$ are the Legendre polynomials. Substitution of Eq. (\ref{eq:mexp}) into Eq. (\ref{eq:Boltzmann}) yields a hierarchy of evolution equations for the multipoles
\begin{subequations}
\label{eq:heem}
\begin{align}
		 & \Psi'_{0} +\frac{qk}{ma} \Psi_{1} = 0 \\ & \Psi'_{1} - \frac{qk}{3ma} (\Psi_{0} - 2\Psi_{2}) = -\frac{mak}{3q} \psi \frac{d\ln f_{0}}{d\ln q} \\  & \Psi'_{l} - \frac{qk}{(2l+1)ma} [l\Psi_{l-1} - (l+1)\Psi_{l+1}] = 0
\end{align}
\end{subequations}

In principle, this is all one needs to study the evolution of linear perturbations of a Newtonian NCDM component. It is, however, convenient to rephrase this as a set of fluid equations for some suitably defined fluid properties. The starting point for this is the expression for the mass density
\begin{align}
\label{eq:Newdensity}
	\rho(1+\delta) &= a^{-3} \int d\Omega \int_{0}^{\infty}d\Pi_{\textrm{New}} \Pi_{\textrm{New}}^2 f_{0} (1+\Psi) m \nonumber \\ &= ma^{-3} \int d\Omega \int_{0}^{\infty}dq q^2 f_{0}(q) (1+\Psi)
\end{align}

This definition of Newtonian energy density is motivated by how it is computed in N-body simulations, i.e. where only the rest mass of particles, as opposed to the total energy $E = \gamma m$, contributes as a source to the gravitational potential. The factor of $(1+3\phi)$, to account for inhomogeneities in the local spatial volume, is also neglected. One can think of choosing a differente gauge, where there are no inhomogeneities in the local spatial volume, e.g. the N-body gauge \cite{1505.04756}. This is a GR correction that will be included in the non-relativistic case.

Substitution of Eq. (\ref{eq:mexp}) into Eq. (\ref{eq:Newdensity}) gives, also defining other relevant fluid properties
\begin{subequations}
\label{eq:Newfp}
\begin{align}
		& \delta \rho = \rho \delta = 4\pi m a^{-3} \int_{0}^{\infty} dq\, q^2 f_{0}(q)  \ \Psi_{0} \\ & \delta P = \frac{4\pi}{3} ma^{-3} \int_{0}^{\infty} dq\, q^2 f_{0}(q)  \Big(\frac{q}{ma}\Big)^2 \Psi_{0} \\ & \rho \theta = 4\pi k a^{-4} \int_{0}^{\infty} dq\, q^2 f_{0}(q) \  q\ \Psi_{1} \\ & \rho \sigma = \frac{8\pi}{3} ma^{-3} \int_{0}^{\infty} dq\, q^2 f_{0}(q)  \Big(\frac{q}{ma}\Big)^2 \Psi_{2}
\end{align} 
\end{subequations}
		
To derive fluid equations, take derivatives of Eq. (\ref{eq:Newfp}) with respect to conformal time, and use Eq. (\ref{eq:heem}) to arrive at
\begin{subequations}
\label{eq:exactfe}
\begin{align}
		& \delta' = -\theta \\ & \theta' = -\mathcal{H} \theta + \frac{\delta P}{\delta \rho} k^2 \delta -k^2 \sigma +k^2 \psi \\ & \sigma' = -2\mathcal{H} \sigma + \frac{4}{15} \Theta -kF_{3}\,,
\end{align}
\end{subequations}
where $\mathcal{H} = a'/a$ and the additional variables $\Theta,F_{3}$ are given by,
\begin{subequations}
\begin{align}
		& \rho F_{3} = \frac{8\pi}{5} ma^{-3} \int_{0}^{\infty} q^2 f_{0}(q) dq \Big(\frac{q}{ma}\Big)^3 \Psi_{3} \\ & \rho \Theta = 4\pi k a^{-4} \int_{0}^{\infty} q^2 f_{0}(q) dq q \Big(\frac{q}{ma}\Big)^2 \Psi_{1} 
\end{align}
\end{subequations}

In order to close the system of Eqs. (\ref{eq:exactfe}), we first write the approximation
\be
	\Psi_{3} \approx \frac{5ma}{qk\tau}\Psi_{2} - \Psi_{1}
\ee

This is a straightforward generalization, suitable to the Newtonian evolution (i.e. let $\epsilon = \sqrt{q^2+a^2m^2} \to am$), of the truncation scheme found in \cite{astro-ph/9506072}. This implies
\be
	kF_{3} \approx \frac{3}{\tau} \sigma - \frac{2}{5} \Theta
\ee

Finally, the approximation for the higher velocity weight fluid properties follows \cite{1104.2935}. This amounts to
\begin{subequations}
\label{eq:fascheme}
\begin{align}	
		& c_{\textrm{eff}}^{2} = \frac{\delta P}{\delta \rho} \approx c_{g}^{2} = \frac{P'}{\rho'} = \frac{5}{3}w = \frac{25}{3} \frac{\zeta(5)}{\zeta(3)} \Big(\frac{T_{0}}{m}\Big)^2(1+z)^2 \\ & \Theta \approx 12 \frac{w}{1+w} c_{g}^{2} \theta 
\end{align}
\end{subequations}
For the sound speed and equation of state, we have used the Newtonian expressions for the background pressure and energy density in Eqs.~(\ref{eq:densitynew}) and ~(\ref{eq:pressurenew}), along with Eq.~(\ref{eq:FD}). Substitution of Eq.~(\ref{eq:fascheme}) into Eq.~(\ref{eq:exactfe}) yields
\begin{subequations}
\label{eq:Newfa}
\begin{align}
		& \delta' = -\theta \\ & \theta' = -\mathcal{H} \theta + c_{g}^{2} k^2 \delta -k^2 \sigma +k^2 \psi \\ & \sigma' = -\Big(2\mathcal{H}+\frac{3}{\tau}\Big) \sigma + 8\frac{w}{1+w} c_{g}^{2} \theta
\end{align}
\end{subequations}

As one can see from Eq. (\ref{eq:Newfa}), this is missing GR corrections, as expected of a Newtonian limit. In this work we concentrate on an SR effect, the shift in the neutrino free-streaming scale, and its impact on the matter power spectrum. Hence, we would like to include GR corrections in our fluid equations. We do so in a self-consistent way by considering the dynamics of a non-relativistic NCDM component, i.e. as follows from Eq.~(\ref{eq:geoNRlimit}) and Eq.~(\ref{eq:GRvelocity}).

The equation of motion is obtained from the action
\begin{align}
\label{eq:NRaction}
	S_{\textrm{NR}} &= \int dt L_{\textrm{NR}}  \nonumber \\ &=  m\int dt \left[\frac{1}{2} a^2 \left(1-2\phi-\psi\right) \left(\frac{d\vec{x}}{dt}\right)^2 - \psi\right]
\end{align}
in the NR limit. This action follows from $-m \int \sqrt{-ds^2}$, after expanding to leading order in metric perturbations and taking the NR limit. The canonical momentum associated to this is
\be
	\vec{\Pi}_{\textrm{NR}} =  \frac{\partial L_{\textrm{NR}}}{\partial \dot{\vec{x}}} = ma^2 \left(1-2\phi-\psi\right) \frac{d\vec{x}}{dt} = (1-\phi)\vec{q}
\ee

The analog of Eq.~(\ref{eq:dynamicsphasespace}) is now, again in the NR limit
\be
\label{eq:dynamicsphasespace2}
	\frac{d\vec{q}}{d\tau} = -ma\vec{\nabla} \psi + \phi' \vec{q} 
\ee

The rest of the procedure to derive fluid equations follows closely what is done in the Newtonian case. The only difference is that we now include the factor of $(1+3\phi)$ to account for inhomogeneities in the local spatial volume, in the definition of the mass density. That is,
\begin{align}
\label{eq:Newdensity2}
	\rho(1+\delta) &= (1+3\phi) a^{-3} \int d\Omega \int_{0}^{\infty}d\Pi_{\textrm{NR}} \Pi_{\textrm{NR}}^2 f_{0} (1+\Psi) m \nonumber \\ &= ma^{-3} \int d\Omega \int_{0}^{\infty}dq q^2 f_{0}(q) (1+\Psi)
\end{align}

Note that this is formally identical to Eq.~(\ref{eq:Newdensity}). The only difference in the fluid equations will then come from the additional GR term in the right hand side of Eq.~(\ref{eq:dynamicsphasespace2}), contributing to the evolution equation of the zeroth multipole of the distribution function, and thus we will arrive at 
\begin{subequations}
\label{eq:NRfa}
\begin{align}
		& \delta' = -\theta + 3\phi' \\ & \theta' = -\mathcal{H} \theta + c_{g}^{2} k^2 \delta -k^2 \sigma +k^2 \psi \\ & \sigma' = -\Big(2\mathcal{H}+\frac{3}{\tau}\Big) \sigma + 8\frac{w}{1+w} c_{g}^{2} \theta\,.
\end{align}
\end{subequations}

\bibliography{Neutrinos_in_N-body_simulations.bib}

\end{document}